\documentclass[aps,amssymb,superscriptaddress,preprint,amsmath]{revtex4}
\usepackage{graphicx} 
\begin{document}
\draft
\newcommand{\ii }{{\rm i}}
\newcommand{\cd }{c}
\title{Nonadiabatic Electron Manipulation in Quantum-Dot Arrays }
\author{Keiji Saito}
\affiliation{Department of Applied Physics, School of Engineering \\ 
University of Tokyo, Bunkyo-ku, Tokyo 113-8656, Japan}
\author{Yosuke Kayanuma}
\affiliation{Department of Mathematical Science, Graduate School of Engineering \\
Osaka Prefecture University, Sakai 599-8531, Japan}
\date{\today }

\begin{abstract}
A novel method of coherent  manipulation of  the electron tunneling in quantum-dots 
is proposed, which utilizes the quantum interference in nonadiabatic double-crossing of 
the discrete energy levels. In this method, we need only a smoothly varying gate 
voltage to manipulate electrons, without a sudden switching-on and off. 
A systematic design of a smooth gate-pulse is presented with a simple analytic 
formula to drive the two-level 
electronic state to essentially arbitrary target state, and numerical simulations for 
complete transfer of an electron is shown for a coupled double quantum-dots 
and an array of quantum-dots. Estimation of the manipulation-time shows that 
the present method can be employed in realistic quantum-dots.

\noindent
\pacs PACS numbers: 73.63.Kv,72.23.Hk,03.67.Lx 
\end{abstract}
\maketitle
\noindent
%\begin{multicols}{2}
%\narrowtext

Quantum-dots (QDs) have been attracting much attention recently because of the 
fundamental interest in the quantum mechanical properties they show as man-made 
atoms and molecules\cite{T_review}. The potential application to future quantum devices 
also stimulates the research interest in QDs. 
Especially, the recent success in the observation of a coherent oscillation of 
an electron in 
double quantum-dots (DQDs) of semiconductors\cite{Hayashi} enhanced the motivation 
to utilize 
QDs as elements 
of new information processing devices (i.e. qubits) based on the principle of 
quantum mechanics\cite{LD98}.
Similar phenomena have been reported also for Josephson
qubits\cite{Josephson}. 
From a theoretical side, 
a number of proposal to control the 
tunneling coherence by applying time-dependent external fields have been presented, 
both for a double-well potential model\cite{Lin,Grossmann,Holthouse} and for a 
simplified two-level model\cite{GH92,YK,Garraway,Miyashita}. See, Grifoni 
and H\"anggi\cite{Grifoni}
for a review on driven tunneling systems.
%See Grifoni and H\"anggi\cite{Grifoni} for a review on driven tunneling systems.
\par
The coherent manipulation of the 
electronic states is a prerequisite 
foundation for the realization of such quantum devices. Therefore, 
it is quite crucial 
to develop techniques how to drive electronic states within the framework of 
quantum mechanics.
The experimental realization of coherent manipulations of electronic 
states in QDs\cite{Hayashi} and Josephson qubits\cite{Josephson} has utilized the 
Rabi oscillation\cite{Rabi} induced by a sudden switching of the gate voltage. For 
example, in order to transfer an excess electron from the left dot to the right dot, 
one applies a rectangular voltage pulse that brings the two-level system to the 
resonant condition suddenly, and after a half period of the Rabi oscillation, 
brings back to the off-resonant state suddenly again\cite{Hayashi}. 
This requires high frequency components in the 
gate fields, which in some cases may pose a difficulty in actual applications.
\par
In the present work, we propose an alternative technique for the coherent manipulation. 
The process requires only a smoothly varying gate voltage in contrast to the 
manipulation by the Rabi oscillation. This is based on the nonadiabatic transition in 
level crossing systems. Therefore, we call it the nonadiabatic manipulation. 
It will be shown that, using a simple analytic formula, we can design a smooth 
temporal profile of the gate voltage that brings about desired transformation of 
the electronic state in coupled QDs. Our proposal here will extend 
the option for the coherent manipulation.
\par
%%%%%%%%%%%%%%%%%%%%%%%%%%%%%%%%%%%%%%%%%%%%%%%%%%%%%%%%%%%%%%%%%%%%
The principle of nonadiabatic manipulation is based on the celebrated 
Landau-Zener formula\cite{Landau,Zener}.
The Landau-Zener mechanism has been recently studied as a possible 
implementation of qubit operations \cite{LZQC1,LZQC2,LZQC3}.
With a suitable sequence of level crossing, interference effects and 
dissipative effects can be used to implement quantum memory 
devices \cite{LZQC2}.
Interference patterns of qubits in a periodic controlling is useful to 
estimate a decoherence time of qubits \cite{LZQC3}.
In the Landau-Zener mechanism, not only a transition probability but also 
a quantum phase is very important and useful for manipulation. 
In the present work, we use an asymptotically exact time evolution of
the wave function to control a transition probability and a quantum phase.
A new aspect of controlling them is to use the transfer matrix technique.
The essence of the transfer matrix was derived by Zener\cite{Zener} 
as early as in 1932, and its usefulness in elucidating the role of
quantum coherence was shown in a periodically driven two-level system\cite{YK}. 
In the followings, we show some examples of nonadiabatic 
manipulation for a DQD and for an array of QDs.
%%%%%%%%%%%%%%%%%%%%%%%%%%%%%%%%%%%%%%%%%%%%%%%%%%%%%%%%%%%%%%%%%%%%%%
\par
First let us consider electronic states in a simple DQD, the Hamiltonian of which 
is given by
\begin{equation}
{{\cal H}(t)} =\varepsilon_{1} (t) |1\rangle \langle 1| + \varepsilon_2 
|2\rangle \langle 2|+ \gamma (|1\rangle \langle 2| + 
|2\rangle \langle 1|),
\label{hamil}
\end{equation}
where $|1\rangle$ and $|2\rangle$ describe the state in which the electron 
occupies a discrete level in the left dot and the right dot, respectively. 
Without loss of generality, we can assume that only the energy of the state $|1\rangle$ 
is modulated by the gate voltage, since only the relative energy is relevant 
for the coherent dynamics. Let us take a gate voltage that drives $\varepsilon_1(t)$ 
from $t=0$ to $t=t_f$ as shown in Fig.1: namely, the diabatic energy 
$\varepsilon_1(t)$ crosses $\varepsilon_2$ twice at $t_1$ and $t_2$. Now we ask what the state vector 
$|\psi(t_f)\rangle$ for the two-level system is 
at the final state of the double-crossing 
provided it starts from $|\psi(0)\rangle$. If the magnitude of the tunneling parameter 
$\gamma$ is small enough as is usually the case, the transition is 
localized around the level crossings. In such a situation, the 
time-evolution of the two-level system can well be decomposed into a coherent succession 
of free propagations and the Landau-Zener type impulsive transitions at around 
level-crossing times $t_1$ and $t_2$. 
\par
We expand $|\psi(t)\rangle$ as
\begin{equation}
|\psi(t)\rangle = C_1(t)|1\rangle +C_2(t)|2\rangle ,
\end{equation}
and define a column vector ${\bf C}(t)\equiv (C_1(t), C_2(t))^T$ where $T$ means 
a transpose. Then 
within the approximation of impulsive transitions, we have 
\begin{equation}
{\bf C}(t_f)=U(t_f,t_2)T(t_2,t_1)U(t_1,0){\bf C}(0),
\end{equation}
where $U(t'',t')$ is a free propagator,
\begin{equation}
U(t'' , t' ) = \left(
\begin{array}{cc}
e^{-\frac{i}{\hbar}\int_{t'}^{t''} du E_{+}(u)  }
& 0 \\
0 & e^{ -{i\over \hbar} \int_{t'}^{t''} du E_{-}(u)  } 
\end{array}
\right),
\end{equation}
in which $E_{+}(u)$ and $E_{-}(u)$ are the adiabatic eigenvalues for the upper and 
the lower branch, respectively, and are given by
$
E_{\pm}=\bigl\{\varepsilon_1(u)+\varepsilon_2 \pm\sqrt{
(\varepsilon_1(u)-\varepsilon_1)^2+4\gamma^2}\bigr\}/2.
$
The matrix $T(t_2,t_1)$ represents the scattering by the double-crossing. The 
time-dependence of $\varepsilon_1(t)$ at the crossings is approximated by a 
linear function with the rate of change $v=|d(\varepsilon_1(t)-\varepsilon_2)/dt|$ 
measured at 
$t=t_1$ and $t_2$. In the present work, we assume for simplicity 
that $v$ is the same at $t_1$ and $t_2$, although this is by 
no means a restriction. Then we find
\begin{equation}
T(t_2,t_1)=M^T V(t_2,t_1) M,
\end{equation}
in which $M$ is the transfer matrix
\begin{equation}
M= \left(
\begin{array}{cc}
\sqrt{q} & -\sqrt{1 - q} e^{i \phi} \\
\sqrt{1 - q} e^{-i\phi} & \sqrt{q} 
\end{array}
\right) ,
\end{equation}
where $q\equiv \exp(-2\pi\delta)$ with $\delta\equiv \gamma^2/\hbar v$, and $\phi$ is the 
Stokes phase,
\begin{equation}
\phi = \pi/4 + {\rm arg}\Gamma (1-i\delta) + \delta({\rm ln}\delta -1 ),
\label{Stokes}
\end{equation}
with the Gamma function $\Gamma (z)$. The propagator $V(t_2,t_1)$ is given by
\begin{equation}
V(t_2 ,t_1 ) = \left(
\begin{array}{cc}
e^{-{i\over \hbar}\int_{t_1}^{t_2} du E_{-}(u)  }
& 0 \\
0 & e^{ -{i\over \hbar} \int_{t_1}^{t_2} du E_{+}(u)  } 
\end{array}
\right).
\end{equation}
We find easily 
\begin{equation}
T(t_2,t_1)=e^{-iS_1}K,
\end{equation}
where $S_1={1\over 2\hbar}\int_{t_1}^{t_2}du \{E_{+}(u)+ E_{-}(u)\}$ and
\begin{equation}
K =\left(
\begin{array}{cc}
qe^{i{S\over 2}}+(1-q)e^{-i(2\phi+{S\over 2}) }
& \sqrt{q(1-q)}(e^{-i(\phi+{S\over 2})} - e^{i(\phi+{S\over 2})}) \\
\sqrt{q(1-q)}(e^{-i(\phi+{S\over 2})} - e^{i(\phi+{S\over 2})})  & 
qe^{-i{S\over 2}}+(1-q)e^{i(2\phi+{S\over 2}) } 
\end{array}
\right),
\end{equation}
where $S$ is the relative phase, 
$$S={1\over \hbar}\int_{t_1}^{t_2}
\{E_{+}(u)-E_{-}(u)\}du ,$$ 
which is proportional to the hatched area in Fig.1. 
The matrix $K$ plays an essential role to determine the population dynamics in our 
manipulation, while other factors only determine the relative phase. Note that the 
relative phase is easily controlled by the free propagation.
\par
In order to realize the complete transfer, we tune the speed of passage 
to yield $q=1/2$, i.e. $v=2\pi\gamma^2/\hbar\log 2$. The $K$-matrix is then 
reduced to
\begin{equation}
K =\left(
\begin{array}{cc}
e^{-i\bar\phi}\cos{\Theta\over 2} & -i\sin{\Theta\over 2} \\
-i\sin{\Theta \over 2}& e^{i\bar\phi}\cos{\Theta \over 2}
\end{array}
\right),
\end{equation}
where $\Theta\equiv S+2\bar\phi$ is the phase factor which comes from the relative  
phase $S$ between the double-crossing and the Stokes phase $\bar\phi$. From 
Eq.(7), the Stokes phase is fixed to be $\bar\phi=0.495039\cdots$. If we introduce the Bloch vector $\vec p$ defined for 
the density matrix $\rho\equiv |\psi(t)\rangle\langle \psi(t)|$ and the Pauli 
matrices $\vec \sigma\equiv (\sigma_x,\sigma_y,\sigma_z)$ by 
\begin{equation}
\rho={1\over 2}(1+ \vec p\cdot \vec\sigma),
\end{equation}
the initial vector $\vec p_i=(0, 0, 1)$ which corresponds to 
$|\psi\rangle =|1\rangle$ is 
transformed by $K$ into 
\begin{equation}
\vec p_f = \bigl(\sin\Theta\cos(\bar\phi-{\pi\over 2}),\, 
\sin\Theta\sin(\bar\phi-{\pi\over 2}),\, \cos\Theta\bigr).
\end{equation}
\par
The above formula indicates that we can drive a two-level system essentially to 
any desired state 
starting from the state $|1\rangle$ by controlling the relative  
phase $S$ for the fixed value of $v$. 
If we set $\Theta=\pi ({\rm mod}2\pi)$, we have $\vec p_f=(0, 0, -1)$,  
which corresponds to the complete transfer from $|1\rangle$ to $|2\rangle$;  
the constructive interference between the two transition paths, $|1\rangle \rightarrow 
|1\rangle \rightarrow |2\rangle$ and $|1\rangle\rightarrow 
|2\rangle \rightarrow |2\rangle$ results in the complete transfer. 
If we set, on the other hand, $\Theta=2\pi ({\rm mod}2\pi)$, we have
$\vec p_f= (0, 0, 1)$ which indicates the complete reflection to the initial state 
because of the destructive interference.\cite{note} \par 
Furthermore, if we choose $\Theta=\pi/2 ({\rm mod} 2\pi)$, $K$-matrix is reduced to
\begin{equation}
K ={1\over \sqrt{2}}\left(
\begin{array}{cc}
e^{-i\bar\phi}& e^{-i\pi/2}\\
e^{-i\pi/2}& e^{i\bar\phi}
\end{array}
\right).
\end{equation}
It is an easy matter to see that, by an appropriate change of phase factors for the 
two-level system, the above $K$-matrix is 
transformed into the Hadamard matrix,
$
{1\over \sqrt{2}}\left(
\begin{array}{cc}
1& 1 \\
1& -1
\end{array}
\right).
$
\par
For illustrative examples, we show some numerical simulations of electron 
transfer between coupled arrays of QDs. Our aim is to transfer an 
electron from QD to QD using smoothly varying gate voltages. In Fig.2, 
the time-dependence of the population in the left dot $P_1(t)$ and the right dot 
$P_2(t)$ are shown for the the initial condition $P_1(0)=1$ and $P_2(0)=0$. The 
parameter values are $\epsilon_1(0)=1.0$meV, $\epsilon_2=0$, and $\gamma=0.01$meV. 
The gate voltage is designed to modulate $\epsilon_1(t)$ as $\epsilon(t)=\epsilon_1(0)
+A(\cos (\omega t) -1)$. The two parameters $A$ and $\omega$ are
adjusted to satisfy the 
conditions $v=2\pi\gamma^2/\hbar\log 2$ at $t=t_1, t_2$ and $\Theta = 3\pi$. 
The time-dependence 
of the adiabatic energies is shown in the inset. As shown in Fig.2, the transfer of the 
electron is almost perfect. 
%%%%%%%%%%%%%%%%%%%%%%%%
Error of manipulation can be estimated as the 
deviation of probability $P_{2}(t)$
from unity after one process. We found a quite small error 
$1-P_{2}(t)\sim 0.0011$.
%%%%%%%%%%%%%%%%%%%%%%%%
It should be noted that, although a large 
energy separation is required before and after the double-crossing events in order that 
the Landau-Zener theory works well, and that the stationary population 
in the respective dot is well defined, the energy separation in the intermediate 
state at $t\simeq \pi/\omega$ need not be very large. 
This can be seen in the oscillation 
in $P_1(t)$ with a relatively large amplitude in the intermediate state. 
Thus, we find that our recipe works well beyond our expectation, 
although it is based on the formalism of Zener which is asymptotically exact 
in the limit of scattering 
from $t=-\infty$ to $t=\infty$. We have also ascertained that the electron can be 
transferred from QD to QD one by one for an array of coupled QDs in which 
the energy levels 
are distributed randomly by a suitably designed temporal profile of the local gate 
voltages according to our method.
\par
In contrast to the local control of the gate voltages, we may transfer an electron 
through an array of QDs by a global control using a time-dependent electric field, 
if the energy levels of the QDs are arranged regularly in a specific manner. 
For example, consider a {\it staggered-dots model} for which the Hamiltonian  
without the electric field is given by
\begin{equation}
{\cal H}_0 =\sum_{l=1}^n \varepsilon^{(0)}_l|l\rangle \langle l| + \sum_{l=1}^{n-1}
\gamma\bigl(|l\rangle\langle l+1| + h. c.\bigr),
\end{equation}
where the energy levels are arranged alternately, $\varepsilon^{(0)}_{2l-1}=\Delta$, 
$\varepsilon^{(0)}_{2l}=0$, as shown in Fig.3 (a). If an oscillating 
electric field ${\bf E}(t)$ is applied along the direction of the dot-array, 
the energy of $l$th dot is modulated as
$\varepsilon_l(t)=\varepsilon^{(0)}_l+(l-1)ea E(t)$, 
where $-e$ is the electric charge and $a$ is the separation between the dots, i.e. 
{\it the lattice constant}. 
By setting $E(t)= E_0\sin(\omega t)$, we can attain a succession of 
double-crossing between neighboring QDs (see  Fig.3(a)). Each level undergoes 
a double-crossing with those in the left dot and the right dot once in a period of 
oscillation. The velocity of 
the energy change is given by
 $v=\vline d(\varepsilon_l(t)-\varepsilon_{l+1})
 /dt\,\vline\,=ea\,\vline\, dE(t)/ dt\,\vline\,$ measured at crossing times. We 
have carried out the simulation of the population dynamics for an array of 10 dots 
for parameter values $\gamma=0.01$meV, $\Delta=0.5$\,mev. 
As shown in Fig.3(b), starting from the dot at $l=1$,  
the electron can be 
transferred from QD to QD successfully to the final target-dot after 5  
oscillation of the global field. 
We obtain an error of manipulation as $1-P_{10}(t)\sim 0.018$ after one process.
If the dots-array is isolated, the electron is  
reflected at the right end of the array, and is transferred back   
to the left end, thus repeating a back-and-forth motion driven by the oscillating 
field. It is interesting to note that the direction of the 
transfer depends on the phase of the oscillating field. If we start from the dot at  
say $\l=5$, it moves to the right by the field $E(t)=E_0\sin(\omega t)$, but to the 
left by $E(t)=-E_0\sin(\omega t)$. In this way, we may carry the electron from a  
initial dot to the target dot at will by an appropriate design of the temporal 
profile of the external electric field.
\par
In actual materials, the quantum system is always disturbed by various sources 
of decoherence. First of all, in order that our proposal is realizable, 
the coherence must be maintained during the Landau-Zener transition time 
$\Delta \tau$\cite{YK}. This is given, in the order of magnitude, as 
\begin{equation}
\Delta\tau \simeq {2\gamma \over v}
={\hbar \over \gamma}{\log 2\over \pi}.
\end{equation}
In addition, the coherence time should
be longer than the period between the double crossing $\Delta T \simeq t_2 - t_1$.
This is roughly given for $\Theta=\pi$ as 
\begin{equation}
\Delta T\simeq 2\sqrt{{\hbar S\over v}}=
{2\hbar\over \gamma}\sqrt{{\log 2(\pi-2\bar\phi)\over 2\pi}}.
\end{equation}
If we use the value $\gamma\sim 10\,\mu$eV, $\Delta\tau$ and $\Delta T$ are 
estimated as $\Delta\tau\simeq 1.4\times 10^{-11}$sec and 
$\Delta T\simeq 6.3\times 
10^{-11}$sec, respectively. Thus $\Delta T+\Delta \tau \sim 7.7\times
10^{-11}$sec will be required for the coherent time.
This is of the same order as the manipulation-time by the Rabi 
oscillation $\hbar\pi/2\gamma$, and is
much shorter than the reported decoherence time 
in the order of 10nsec\cite{Hayashi}.
\par
To summarize, we proposed a novel method to manipulate electronic states in quantum 
dots, which utilizes the quantum coherence between transition paths in the Landau-Zener 
type successive level-crossings. Starting from a localized state to one of the levels, 
we can reach arbitrary configuration of the two-level system. This may be regarded as 
a Mach-Zehnder type interference {\it in the time domain}, in which the dynamical phase 
between the crossings plays a role of the optical path-length. We would like to stress 
here that the interference effect between nonadiabatic transition paths in driven 
systems will provide new quantum phenomena, and should be exploited for 
tools of electron manipulation \cite{FKH04}. The coherent destruction of 
tunneling found by Grossmann {\it et al.}\cite{Grossmann},
for example, can be regarded as 
a result of destructive interference between transition paths\cite{YK}.
\par
We thank Professor P. H\"anggi for stimulating discussions. We also thank 
Dr. T. Hayashi and Dr. Y. Hirayama of NTT Basic Research Laboratory for 
valuable discussions on the experimental details. This work was partially supported by 
the Grant-in-Aid for Scientific Research from the Ministry of Education, 
Culture, Sports, Science and Technology.

\newpage
\begin{center}
\begin{figure} 
\includegraphics[height=15cm,width=20cm]{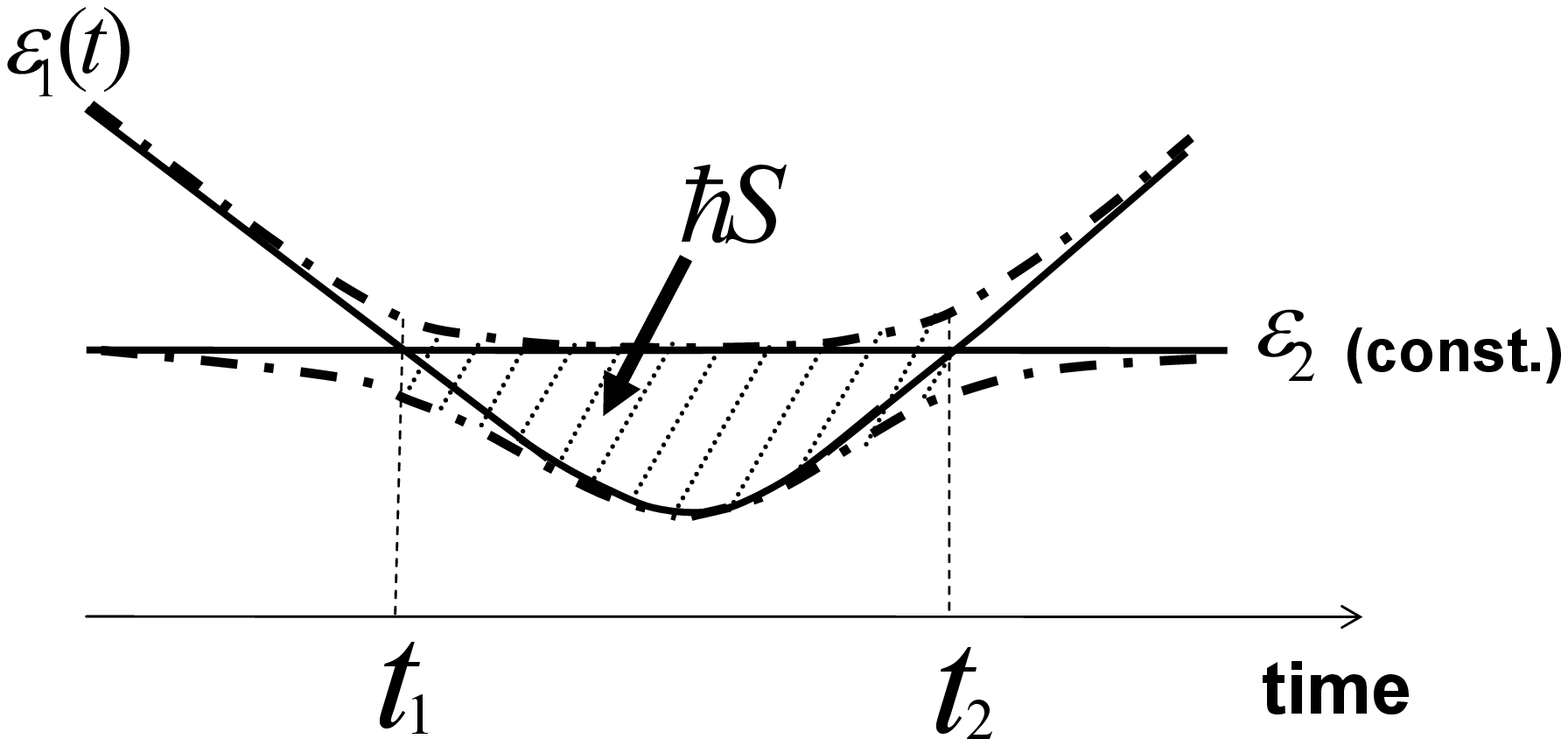} 
\caption{Schematic picture of time-dependence of energies of dots. 
The solid lines are the energies of two dots, and the dashed lines are the
eigenvalues.}
\end{figure} 
\end{center}

\begin{center}
\begin{figure}
\includegraphics[height=20cm,width=30cm]{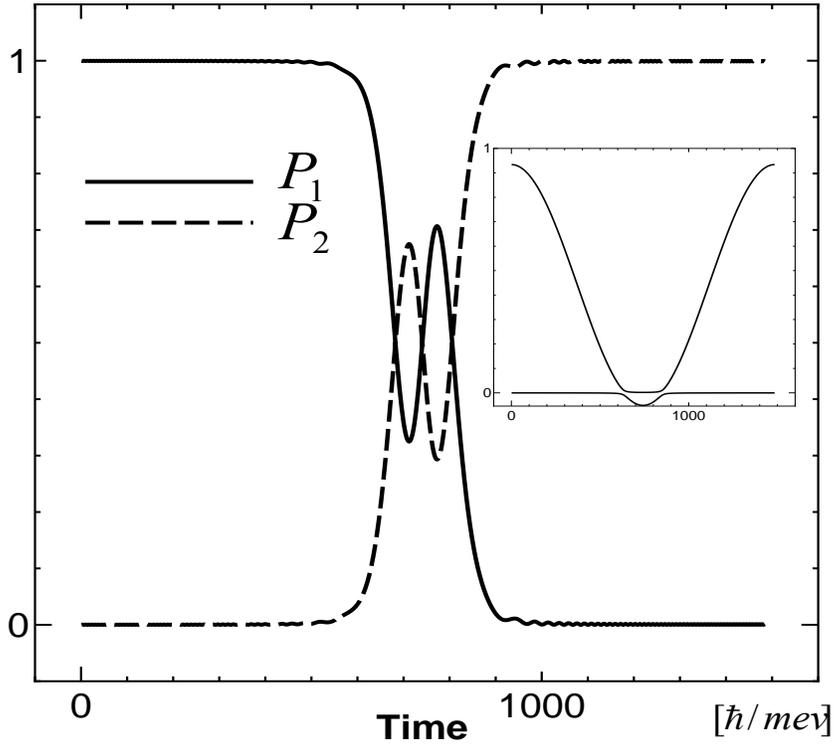} 
\caption{The probabilities $P_{1}(t)$ and $P_{2}(t)$ 
are shown as a function of time. 
In the inset, the time dependences of eigenenergies are also shown.
The unit of time is $\hbar/meV \simeq 4.14\times10^{-13}$seconds.}
\end{figure}
\end{center}
\begin{center}
\begin{figure}
\includegraphics[height=15cm,width=20cm]{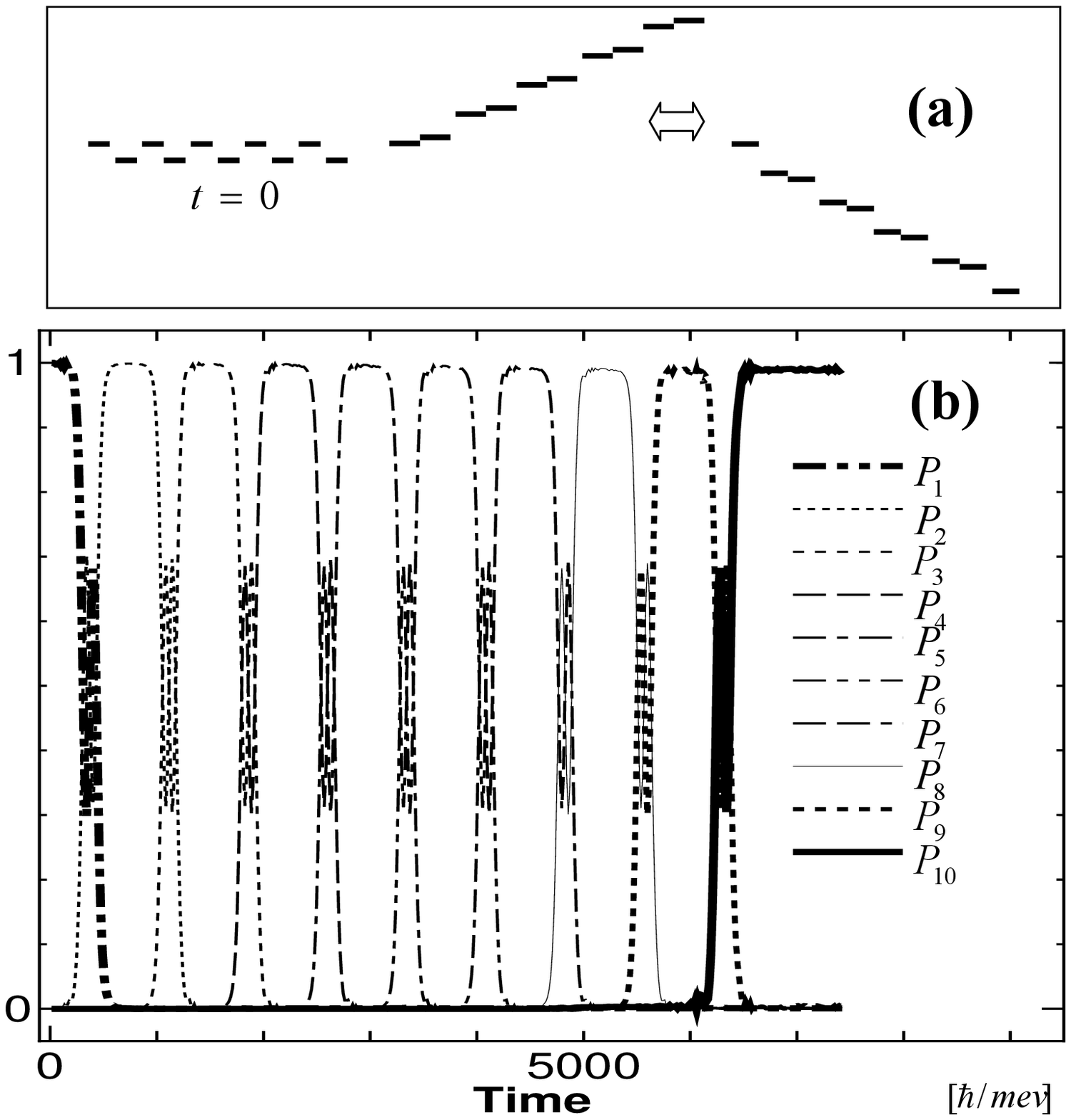} 
\caption{Global Control by an oscillating electric field. 
(a): Schematic picture of time dependence of the staggered dots model. 
(b): The probability finding 
electron in each dot is shown as a function of time.}
\end{figure}
\end{center}

\begin{thebibliography}{notitle}
\bibitem{note} In this case, the state vector $|1\rangle$ is 
transformed into $e^{-i(S_1+\bar\phi-\pi)}|1\rangle$ 
by the scattering matrix $T(t_1,t_2)$. 
Since $-S_1$ is nothing but the nonadiabatic dynamical phase\cite{Aharonov}, 
the additional phase  $\chi_g\equiv \pi-\bar\phi$ 
is the geometrical phase for this specific nonadiabatic cyclic 
evolution\cite{YK2}. 
In fact, it can be 
ascertained that $\chi_g$ is equal to half of the solid angle of the 
trace of $\vec p$ subtended at the origin 
in the Bloch sphere.
\end{thebibliography}
\end{document}